\documentclass[peerreview,10pt,draftcls,onecolumn]{IEEEtran}
\usepackage{graphicx}
\usepackage{cite}

\begin{document}

\title{High-Performance Carbon Nanotube Field-Effect Transistor with Tunable Polarities}

\author{Yu-Ming~Lin,
        Joerg~Appenzeller,
        Joachim~Knoch, and
        Phaedon~Avouris,%
\thanks{Y.-M.~Lin, J.~Appenzeller, and Ph.~Avouris are with the IBM T.\ J.\ Watson Research Center, Yorktown Heights, NY
10598, USA (email: yming, joerga, avouris@us.ibm.com).}%
\thanks{J.~Knoch is with the Institute of Thin Film and Interfaces and Center of Nanoelectronic Systems, Forschungszentrum Julich, D-52454, Julich, Germany.}}
\maketitle
\begin{abstract}
State-of-the-art carbon nanotube field-effect transistors (CNFETs)
behave as Schottky barrier (SB)-modulated transistors. It is known
that vertical scaling of the gate oxide significantly improves the
performance of these devices. However, decreasing the oxide
thickness also results in pronounced ambipolar transistor
characteristics and increased drain leakage currents. Using a
novel device concept, we have fabricated high-performance,
enhancement-mode CNFETs exhibiting $n$ or $p$-type unipolar
behavior, tunable by electrostatic and/or chemical doping, with
excellent OFF-state performance and a steep subthreshold swing ($S
=63$\,mV/dec). The device design allows for aggressive oxide
thickness and gate length scaling while maintaining the desired
device characteristics.
\end{abstract}

\begin{keywords}
carbon nanotube, Schottky barrier, field-effect transistor, doping
\end{keywords}
\IEEEpeerreviewmaketitle

\section{Introduction}

Since the discovery of carbon nanotubes in the early
1990s\cite{Iijima_Nature1993,bethune_nature93}, the field has
witnessed an immense growth. Due to the small diameter ($\sim
1$\,nm), nanotubes are ideal candidates to study one-dimensional
electrical transport phenomena, even at room temperature. In
particular, single-walled carbon nanotubes, which consist of a
single layer of graphene sheet wrapped up to form a seamless
tube\cite{dresselhaus_CNTbook}, possess exceptional electrical
properties, such as high current carrying
capability\cite{yao_PRL2000} and excellent carrier
mobility\cite{Fuhrer_Nano2002}.  Both theory and experiments have
demonstrated that these tubes can be either metallic or
semiconducting, depending on the chirality of their atomic
structure with respect to the tube axis. Using semiconducting
nanotubes, carbon nanotube field effect transistors (CNFETs) have
been first realized and reported in Refs.\cite{martel_APL1998} and
\cite{tans_Nature1998}. These early CNFETs, however, showed poor
device characteristics. Since then, remarkable progress has been
made to improve the performance of CNFETs by approaches such as:
(1) reducing the gate oxide thickness\cite{wind_APL2002}; (2)
adopting high-$\kappa$ dielectrics for the gate
oxide\cite{bachtold_Science2001,appenzeller_PRL2002,javey_Nature2002};
(3) using electrolyte as the gate
\cite{rosenblatt_Nanolett2002,Siddons_Nano2004}; (4) reducing
contact resistance by choosing proper contact
metals\cite{Javey_NanoLett2004} or post-process
treatments\cite{martel_PRL2001}. These techniques have
significantly advanced CNFET devices to exhibit characteristics
rivaling those of state-of-the-art Si-based MOSFETs. With the
scaling limit of conventional CMOS in sight\cite{Frank_IEEE2001},
carbon nanotubes show great promise among the materials
investigated so far for post-Si technology.

Most CNFETs studied so far have adopted a back-gate, top-contact
geometry (see \cite{McEuen_IEEE2002} and \cite{avouris_IEEE2003}
for a review), as shown in Fig.\,\ref{thinoxCNFET}(a).  In this
back-gate configuration, the nanotubes are dispersed or grown on a
conducting substrate covered by an insulating layer. Two metal
contacts are deposited on the nanotube to serve as source and
drain electrodes, while the conducting substrate is the gate
electrode in this three-terminal device. This CNFET structure,
albeit simple, has been extensively studied to provide invaluable
insights and understanding of fundamental properties of nanotubes.
However, these CNFETs are found to possess several unfavorable
features that severely constrain their usefulness in vital
applications such as CMOS-like circuits. In order to address the
disadvantages of a simple back-gated CNFET, in this report we
present a novel device concept for CNFETs. In particular, our
device design is capable of producing pure $n$- and/or $p$-type
CNFETs with improved OFF-state performance and abrupt switching
behavior close to theoretical limits.

First we review important electrical characteristics of the
back-gated CNFET as shown in Fig.\,\ref{thinoxCNFET}(a). The
understanding of  operation principles and physics of this device
is indispensable in order to obtain desired functionality and
further improvements. Figure\,\ref{thinoxCNFET}(b) shows the
subthreshold characteristics, the drain current $I_d$ measured as
a function of gate voltage $V_{\rm gs}$, of a back-gated CNFET for
different drain voltages ($V_{\rm ds}$). For all the measurements
presented in this report, the source electrode is grounded. This
CNFET device is fabricated on a heavily $p$-doped Si substrate
covered by a thin layer (10-nm thick) of thermally-grown SiO$_2$
with titanium (Ti) contacts. In Fig.\,\ref{thinoxCNFET}(b), two
distinct features are observed. First, as the gate voltage
increases from negative values, the current decreases, reaching a
minimum value of $I_{\rm off} = 8\times 10^{-12}, 2\times
10^{-10}$, and $10^{-9}$\,A for $V_{\rm ds}=-0.1$, -0.4, and
-0.7\,V, respectively, and then rises again, exhibiting an
ambipolar transistor characteristic. Second, as the drain bias
becomes more negative, the minimum current $I_{\rm off}$ increases
exponentially and the gate voltage corresponding to this minimum
current shifts to more negative values, showing a strong $V_{\rm
ds}$-dependent $I_{\rm off}$.

In order to understand the results of Fig.\,\ref{thinoxCNFET}(b),
Figures\,\ref{bands2}(a) and (b) depict schematic band diagrams of
a CNFET for negative and positive $V_{\rm gs}$, respectively. We
note that, different from a conventional MOSFET, switching of a
CNFET is dominated by the modulation of Schottky barriers (SBs)
formed at the nanotube/metal
contacts\cite{freitag_APL2001,martel_PRL2001,appenzeller_PRL2002,heinze_PRL2002}.
As indicated in Figs.\,\ref{bands2}(a) and (b), at sufficiently
negative and positive gate voltages, Schottky barriers are
effectively thinned to enable hole injection ($V_{\rm gs}<0$) from
the source and electron injection ($V_{\rm gs}>0$) from the drain
contact into the nanotube, respectively. Compared to the tunneling
phenomena in a 3D bulk device, this tunneling current in CNFETs
can be quite significant because a barrier width ($\sim$ depletion
length $L_D$) as small as a couple of nanometers can be obtained
for a ultra-thin body object, such as a nanotube, when thin gate
dielectrics are used\cite{appenzeller_PRL2004}. We also notice
that in Fig.\,\ref{bands2}, there is little band bending in the
nanotube body because the carrier mean-free-path in nanotubes can
be as long as $\sim 500$\,nm even at room
temperature~\cite{wind_PRL2003,javey_Nature2003}, so that carrier
transport is nearly ballistic for the nanotube channel length
($\sim 300$\,nm) considered here. In Fig.\,\ref{thinoxCNFET}(b),
the left and right current branches with respect to the minimum
current $I_{\rm off}$ are therefore contributions by hole and
electron currents, and are referred to in the following as $p$-
and $n$-branches, respectively. At the minimum current $I_{\rm
off}$, the hole current is equal to the electron current. Since
the electron current is due to the tunneling at the drain contact,
the $n$-branch exhibits a much stronger $V_{\rm ds}$ dependence
than the $p$-branch (see Fig.\,\ref{bands2} for two different
$V_{\rm ds}$), resulting in a $V_{\rm gs}$ shift of the $n$-branch
as a function of $V_{\rm ds}$ and a strongly $V_{\rm
ds}$-dependent $I_{\rm off}$ as shown in
Fig.\,\ref{thinoxCNFET}(b). All these electrical characteristics
of CNFETs have been studied previously, and detailed discussions
can be found in
Refs.\cite{heinze_PRL2002,marko_APL2003,Guo_IEEE2004}.

Both simulations and experimental results have shown that the
switching performance of CNFETs, measured by the inverse
subthreshold slope $S=(d\log I_d/dV_{\rm gs})^{-1}$, and the
ON-state current can be improved by decreasing the gate oxide
thickness\cite{wind_APL2002,appenzeller_PRL2002,heinze_PRB2003,javey_Nature2002,bachtold_Science2001}.
However, the enhanced gate control in a CNFET always also results
in a higher $I_{\rm off}$ with a more pronounced $V_{\rm ds}$
dependence as explained before\cite{marko_APL2003}. In addition,
from application's point of view, both $n$ and $p$-type
transistors with unipolar characteristics are required for
CMOS-like logic circuits, while thin-oxide CNFETs are usually
ambipolar, as shown in Fig.\,\ref{thinoxCNFET}(b). In order to
advance nanotubes for future transistor technology, these
disadvantages associated with vertical scaling have to be
overcome. We have recently demonstrated the conversion of an
ambipolar CNFET to a unipolar transistor using a partially-gated
structure where a trench is created near one of the contacts to
eliminate drain leakage\cite{lin_Nanolett04}. However, it was also
found that this partially-gated structure results in a reduced
ON-current and is not suitable for scaling to small lateral
lengths due to fringing field impacts.

In this report, we present a novel device concept for CNFETs using
dual gates to eliminate ambipolar characteristics and create pure
$n$- and/or $p$-type devices with excellent OFF-state performance
and a steep subthreshold swing. This is achieved by a combination
of electrostatic and/or chemical doping along the tube channel. By
introducing a distinct $p/i/p$ or $n/i/n$ band structure  along
the nanotube (where $n,p$, and $i$ represent an $n$-doped, a
$p$-doped, and an undoped (intrinsic) nanotube segment,
respectively), we have successfully modified the switching
mechanism of our CNFET\cite{Lin_DRC2004}. Instead of contact
switching by modulating Schottky barriers, our device operates by
varying the band potential in the middle portion of the nanotube
body, called bulk-switching, and achieves an inverse subthreshold
slope $S\sim 63$\,mV/dec, the smallest value reported for CNFETs
to date. Furthermore, this unique device structure affords
aggressive gate length scaling without deteriorated device
performance.

\section{Novel Dual-Gate CNFET Design}

Figure\,\ref{dblGate} shows the SEM image as well as the device
cross-section of a CNFET with the proposed dual-gate
structure\cite{Lin_DRC2004}. Different from the back-gated CNFET
shown in Fig.\,\ref{thinoxCNFET}(a), the dual-gate CNFET possesses
an additional Al gate electrode placed underneath the nanotube
between the source and drain contacts. To fabricate the dual-gate
CNFET, the Al gate (20-nm thick and  $\sim 200$-nm wide) is first
deposited on a $p$-doped Si substrate covered with 10-nm-thick
SiO$_2$, followed by the oxidation of Al metal in moisturized
oxygen at $\sim 160^{\circ}$C for 1.5 hours to form a thin layer
of Al$_2$O$_3$ ($\kappa\simeq 5$).  Both C-V and ellipsometry
measurements of the Al$_2$O$_3$ layer indicate an oxide thickness
of about 4\,nm, in agreement to the result reported in Ref.\,
\cite{bachtold_Science2001}. Nanotubes produced by laser
ablation~\cite{thess_Nature1996} with an average diameter of $\sim
1.4$\,nm are then spun onto the substrate from solution. Finally,
source/drain contacts made of Ti are deposited on the nanotube,
each with a spacing of $\sim 200$\,nm relative to the Al gate. For
the following discussions, the Al gate region is denoted as region
B, and the area between the Al gate and the source/drain contacts
are denoted as regions A (see Fig.\,\ref{dblGate}).

In our design, the Al gate is the primary gate that governs the
electrostatics and the switching of the nanotube bulk channel in
region B, while the Schottky barriers at the nanotube/metal
contacts are controlled by the Si back gate (substrate), which
also prevents the electrostatics in region A from being influenced
by the Al gate.  It was previously observed that carrier injection
at the contacts may still be modulated by a middle gate due to
fringing field impacts\cite{javey_Nature2002,lin_Nanolett04}. To
verify that the Schottky barriers at the contacts are not affected
by the Al gate voltages in our design, Figure\,\ref{BGfloating}
shows the measured current as a function of Al gate voltage
$V_{\rm gs-Al}$ when the Si back gate $V_{\rm gs-Si}$ is kept
floating. We find that our device is always OFF with a low current
in the noise level ($\le$\,pA), independent of $V_{\rm gs-Al}$,
indicating an insignificant impact of the Al gate fields at the
contact regions. The result in Fig.\,\ref{BGfloating} implies that
our CNFET can operate as a bulk-switched transistor rather than an
SB-CNFET since the switching using the Al gate does not impact the
contact Schottky barriers.

CNFETs with potential profiles similar to our device have been
previously fabricated by using one or more middle gates on top of
the nanotube\cite{wind_PRL2003,javey_Nature2002,Javey_nanoLett04}.
In those CNFETs, the nanotube is sandwiched between the top
gate(s) and the back gate, and the back gate field still
contributes to the electrostatics inside the nanotube, resulting
in a lower efficiency of the top gate when the back gate is kept
at any constant voltage\cite{footnote1}. In contrast, the nanotube
potential of region B in our design is exclusively determined by
the Al gate because the Al metal layer screens the field from the
Si gate, giving rise to an ideal switching behavior. Another
important advantage of our design is the possibility of
introducing a well-defined chemical doping profile along the
nanotube to further tailor transistor characteristics (see
Sec.\,\ref{sec:chemical}), while no chemical doping profile can be
easily introduced in a completely top-gated CNFET.

\subsection{Electrostatic Doping Effect}

In this section we present electrical characteristics of our
dual-gate CNFETs at various gate voltage ($V_{\rm gs-Al}$ and
$V_{\rm gs-Si}$) configurations.  In particular, we utilize
electrostatic doping effects in CNFETs to eliminate ambipolar
characteristics in an SB-CNFET and to obtain a bulk-switched
transistor possessing a tunable polarity ($n$ or $p$), steep
subthreshold swing, excellent OFF-state performance, and
suppressed $V_{\rm ds}$ dependence\cite{footnote5}. This is
achieved by a distinct $n/i/n$ or $p/i/p$ band profile in our
dual-gate CNFET.

Figure\,\ref{npBranch}(a) shows the subthreshold swing of a
dual-gate CNFET at $V_{ds}=-0.6$\,V when the two gate voltages are
equal ($V_{\rm gs-Si}=V_{\rm gs-Al}$). In this gate voltage
configuration, the device is essentially equivalent to a standard
back-gated CNFET as shown in Fig.\,\ref{thinoxCNFET}(a), and the
switching behavior is mainly determined by the Schottky barriers
at the contacts, resulting in the expected ambipolar behavior. In
Fig.\,\ref{npBranch}(a), the OFF-current $I_{\rm off}$ is  below
our instrument sensitivity level ($10^{-13}$\,A)\cite{footnote4}.
The dual-gate CNFET possesses drastically different subthreshold
characteristics ($I_d$ measured as a function of $V_{\rm gs-Al}$)
when the Si gate voltage $V_{\rm gs-Si}$ is kept constant.  As
shown in Fig.\,\ref{npBranch}(b), the same dual-gate CNFET
exhibits clear $p$ and $n$-type unipolar properties for $V_{\rm
gs-Si}=-2$\,V ($\bullet$) and +1.6\,V ($\circ$), respectively. In
Fig.\,\ref{npBranch}(b), positive (+0.6\,V) and negative (-0.6\,V)
drain voltages are used for $n$-type and $p$-type branches,
respectively, in order to illustrate regular $n$ and $p$-FET
device operations. This $p$-FET ($V_{\rm gs-Si}<0$) and $n$-FET
($V_{\rm gs-Si}>0$) behavior of the dual-gate CNFET can be
understood by the schematic band diagrams shown in
Figs.\,\ref{bands3}(a) and (b), respectively.  For a sufficiently
negative (or positive) Si gate voltage, the Schottky barriers are
thinned enough to allow for hole (or electron) tunneling from the
metal contacts into the nanotube channel, and regions A become
electrostatically doped as $p$-type (or $n$-type), resulting in a
$p/i/p$ (or $n/i/n$) band profile that allows only hole (or
electron) transport in the nanotube channel.
 The dual-gate CNFET is switched
ON and OFF by varying the Al gate voltage that alters the barrier
height for carrier transport across region B. In this
configuration, regions A serve as extended source and drain, and
our device operates in a fashion similar to a conventional MOSFET
through bulk-switching in region B. We note that for Si MOSFETs,
transistors with electrically-induced source/drain regions have
been fabricated by using additional side
gates~\cite{Sone_EDL1998}, and our dual-gate structure is
reminiscent of the straddle-gate Si MOSFETs demonstrated by Tiwari
et al~\cite{tiari_IEDM98}.

The results in Fig.\,\ref{npBranch}(b) are in drastic contrast to
the ambipolar behavior of the same device shown in
Fig.\,\ref{npBranch}(a), proving the successful elimination of
ambipolar characteristics and the first experimental demonstration
of controlling the CNFET polarity by means of electrostatic
effects. It is interesting to note the $n$-FET branch in
Fig.\,\ref{npBranch}(b) exhibits a non-monotonic behavior with the
current rising for $V_{\rm gs-Al}\le -1$\,V, as indicated by the
circle in Fig.\,\ref{npBranch}(b). This feature is due to
band-to-band tunneling that occurs when the two gate voltages are
different enough such that the valence band in region B is higher
than the conduction band in regions A (or vice versa). Further
discussion of this band-to-band tunneling phenomenon is reported
elsewhere\cite{appenzeller_B2B}.

In our dual-gate CNFETs, the Si back gate plays an important role
in determining the type of majority carrier and the device
ON-current. In order to study the impact of the Si gate voltage on
the device performance, Figure\,\ref{variousVBG}(a) shows the
subthreshold characteristics of a dual-gate CNFET for different Si
back gate voltages $V_{\rm gs-Si}$, with particular emphasis on
the $p$-FET branches ($V_{\rm gs-Si}, V_{\rm ds}<0$). We first
note that for $V_{\rm gs-Si} > -1.0$\,V, the ON-current is
relatively low because the current is limited by the carrier
injection at the Schottky barriers rather than the barrier in
region B. For $V_{\rm gs-Si} = -0.5$\,V, in particular, $I_d$ is
independent of $V_{\rm gs-Al}$  and the device is essentially OFF
for the entire $V_{\rm gs-Al}$ range, indicating the Schottky
barrier at the contact prohibits carrier emission for this Si gate
voltage. As $V_{\rm gs-Si}$ becomes more negative, the Schottky
barrier becomes thinner to facilitate hole injection, giving rise
to a higher ON-current. At sufficiently negative $V_{\rm gs-Si}$
($\le -2$\,V), the device operates as an enhancement-mode $p$-FET
with a negative threshold voltage and an inverse subthreshold
slope $S\sim 63$\,mV/dec. We note that this $S$ value is the
smallest slope reported for CNFETs to date, and is close to the
theoretical minimum value of $(k_BT/q)\ln 10 \simeq 60$\,mV/dec
attainable for conventional MOSFETs at room temperature (where
$k_B$, $T$, and $q$ are the Boltzmann constant, temperature, and
electron charge, respectively).  In comparison, the same device
possesses $S\sim 100$\,mV/dec when operated as an SB-CNFET with
$V_{\rm gs-Al}=V_{\rm gs-Si}$ (see Fig.\,\ref{variousVBG}(b)).
The results indicate that in addition to the elimination of
ambipolar characteristics, a steeper subthreshold swing can also
be obtained in our devices due to bulk-switching enabled by the
distinct dual-gate design. The nearly-ideal $S$ value (63\,mV/dec)
measured for our dual-gate CNFET suggests a gate efficiency close
to 1, which is made possible by the dual back gate design that
allows exclusive Al gate control in region B. We further note that
in addition to the factors that affect the inverse subthreshold
slope $S$ for conventional MOSFETs, the $S$ value of our dual-gate
CNFETs also depends on tunneling properties at the Schottky
barriers, as explained in the following.

In the vicinity of the Schottky barriers in regions A, the carrier
energy distribution $\rho(E)$ is proportional to the product of
$T(E)\cdot f(E,T)$, where $T(E)$ is the tunneling probability
through the Schottky barrier and $f(E,T)$ is the Fermi-Dirac
distribution in the metal at temperature $T$. As carriers travel
away from the contact, inelastic scattering processes such as
electron-phonon interaction will thermalize the carrier
distribution back to the equilibrium distribution characterized by
the ambient temperature $T$, which is the important parameter in
deriving the theoretical value of $S=(k_BT/q)\ln 10$. However, at
low fields, the carrier mean-free-path in CNFETs can be as long as
half a micrometer and the transport is essentially ballistic over
the entire region A ($\sim 200$\,nm). Without an effective
mechanism \cite{footnote7} to relax the non-equilibrium carrier
energy distribution, the carriers arriving at the barrier in
region B possess the same distribution as $\rho(E)$, which can, in
principle, deviate significantly from a well-tempered Fermi-Dirac
distribution if the tunneling probability $T(E)$ is highly
energy-dependent. Since tunneling through a Schottky barrier
always favors higher energy carriers, the effective temperature
$T_{\rm eff}$ of the carrier distribution $\rho(E)$ is higher than
the lattice temperature $T$\cite{Freitag_NanoLett04}, causing a
larger $S$ value than $(k_BT/q)\ln 10$. Therefore, from the aspect
of device performance, a Schottky barrier with $T(E)\sim$ constant
($\sim 1$) is important not only for a high ON-current, but also
essential to obtain an inverse subthreshold slope close to
60\,mV/dec at room temperature.

We have also studied the $V_{\rm ds}$ dependence of our device
performance, as shown in Fig.\,\ref{variousVds} for the measured
subthreshold characteristics of a dual-gate CNFET at different
drain voltages. In Fig.\,\ref{variousVds}, the Si back gate
$V_{\rm gs-Si}$ is always kept at -4\,V, and  the device exhibits
$p$-FET characteristics. Compared to the results in
Fig.\,\ref{thinoxCNFET}(b), the subthreshold swing of the
dual-gate CNFET exhibits a much weaker dependence on the drain
voltage $V_{\rm ds}$. The dual-gate CNFET possesses a very low
$I_{\rm off}<100$\,fA, and $I_{\rm off}$ is almost independent of
$V_{\rm ds}$. This dramatic improvement of OFF-state performance
for dual-gate CNFETs is not unexpected because the minority
carrier injection from the drain electrode is effectively
eliminated by the distinct $p/i/p$ (or $n/i/n$) band profile in
our design (see Fig.\,\ref{bands3}).

By using a novel dual-gate structure, we have successfully
modified the switching mechanism of a CNFET from Schottky barrier
modulation to bulk-channel switching, resulting in a unipolar
transistor with tunable polarity, excellent subthreshold slopes,
and drastically improved OFF-state.

\subsection{Tuning via Chemical Doping}
\label{sec:chemical} In the previous section, we have demonstrated
CNFETs with improved performance due to a distinct $p/i/p$ (or
$n/i/n$) potential profile along the nanotube by means of
electrostatic doping in selected areas (i.e.\ regions A). By
varying the Si back gate voltage, the polarity of the dual-gate
CNFET can be readily tuned. The role of the Si back gate, in
principle, can be replaced by introducing proper chemical dopants
into regions A in order to achieve the same band bending profile.
The use of chemical dopants in regions A is attractive because the
device can be operated with the back gate grounded ($V_{\rm gs-Si}
= 0$)
or ultimately without the presence of any back gate, which is
essential for high frequency applications to minimize parasitic
capacitances. Controlled $n$-type chemical doping of nanotubes has
been accomplished previously by using alkali metals or air-stable
polymers to create $n$-type
transistors\cite{derycke_NanoLett2001,Shim_JACS2001,bockrath_PRB2000,kong_APL2000,marko_APL2004,jia_DRC2004}
and $p/n/p$ devices\cite{kong_APL2002}. However, in those devices,
gating always occurs over the entire nanotube.  Based on the
dual-gate CNFET design, next we discuss our results on chemical
doping in conjunction with electrostatic doping. In particular,
$n$-type chemical dopants are used here to obtain an $n/i/n$
doping profile along the nanotube.

In order to obtain a chemical doping profile along the tube, a
resist layer consisting of HSQ is patterned on top of the Al gate
area of a dual-gate CNFET. HSQ (Hydrogen SilsesQuioxane) is a
negative resist for e-beam lithography, and is compatible with our
chemical doping processes for nanotubes. Figure\,\ref{dblGate_HSQ}
shows the SEM image and the schematic cross-section of a dual-gate
CNFET patterned with a layer of HSQ resist. The HSQ layer protects
the nanotube segment (region B' in Fig.\,\ref{dblGate_HSQ}) from
being influenced by chemical dopants, so that a $p/i/p$ (or
$n/i/n$) chemical doping profile can be obtained with respect to
regions A' and B'. In our design, the Al gate and the HSQ layer
permit independent electrostatic and chemical doping profiles
along the tube channel.

 After the HSQ patterning, Figure\,\ref{p2n}(a) shows measured $I_d$ vs.\
$V_{\rm gs-Al}$ of a dual-gate CNFET (Ti/Pd contact) at different
$V_{\rm gs-Si}$ prior to any chemical doping
treatment\cite{foornote6}. We note that this device operates as a
$p$-FET even for $V_{\rm gs-Si}\ge 0$, suggesting that this CNFET
was unintentionally $p$-doped during the fabrication
process\cite{javey_Nature2002}. However, Figure\,\ref{p2n}(a) also
points out an important fact which is that the nanotube segment in
region B' is not subject to this unintentional $p$-doping because
the CNFET is still in the OFF state at $V_{\rm gs-Al}=0$\,V.
Therefore, this CNFET possesses a $p/i/p$ doping profile due to
the unintentional $p$-doping effect after the HSQ patterning.  The
device is then dipped in an ethanol solution containing 25wt\% of
PEI for 4 hours, followed by an ethanol rinse. Here we use PEI
(polyethylene imine), an $n$-type dopant for
nanotubes\cite{Shim_JACS2001}, to demonstrate chemical doping
effects in our devices. Figure\,\ref{p2n}(b) shows the measured
subthreshold characteristics of the same CNFET after PEI doping,
exhibiting a drastically different behavior from that of the
original device shown in Fig.\,\ref{p2n}(a). The clear $n$-FET
characteristics for $V_{\rm gs-Si} \ge -0.5$\,V shown in
Fig.\,\ref{p2n}(b) indicate successful $p$-FET to $n$-FET
conversion by doping regions A' with PEI. One important feature
observed in Figs.\,\ref{p2n}(a) and (b) is that both $p$-FET and
$n$-FET devices before and after PEI doping operate as
enhancement-mode transistors with threshold voltages $V_{\rm th}$
of about -0.7\,V and +0.4 V, respectively. This is due to the
$p/i/p$ and $n/i/n$ chemical doping profile enabled by the HSQ
resist layer. Without our distinct doping profile, the threshold
voltage would shift with increasing doping level, resulting in a
depletion-mode transistor at high doping
levels\cite{marko_APL2004}.

The operation of the dual-gate CNFET with a chemical doping
profile can be understood by the band structures depicted in
Fig.\,\ref{doped_band3}. We note that since the area covered by
HSQ resist (region B') is narrower than the Al gate (region B),
there exists an additional band bending within region B where both
the Al gate and the chemical dopants are effective. This
additional step in the potential profile has little effect on the
transistor performance because the switching behavior is mainly
determined by the total barrier height rather than the detail
structure of the barrier. Band-to-band tunneling currents, on the
other hand, may be substantially suppressed because of this
step-like band structure in region B.

\section{Simulation and Discussion}

Last, we present simulation results for our dual-gate CNFETs. In
the simulation, we consider a CNFET consisting of a nanotube
connected to two semi-infinite source/drain metallic contacts,
forming Schottky barriers. The dual-gate structure is explicitly
taken into account by three gate segments between the contacts,
and is characterized by lengths $L_A$ and $L_B$ for regions A and
B, respectively. The charge in and the current through the CNFET
is calculated self-consistently using the non-equilibrium Green's
function formalism together with a modified 1D Possion equation. A
quadratic dispersion relation is assumed in the conduction and
valence band. Further details regarding the simulation can be
found elsewhere\cite{knoch_Solid04}.

Figure\,\ref{exp_sim} shows both measured and simulated
subthreshold characteristics of a dual-gate CNFET (without
chemical doping) at different Si back gate
voltages\cite{footnote2}. In the simulation, we assumed a nanotube
energy gap $E_g$ of 1.0\,eV\cite{footnote3} and a Schottky barrier
height $\Phi_B=0.3$\,eV for holes. In Fig.\,\ref{exp_sim}, the
simulated current $I_d$ as a function of Al gate voltages $V_{\rm
gs-Al}$ exhibits excellent agreement with measured data over a
wide range of $V_{\rm gs-Al}$ for various Si back gate voltages.
In particular, both experiment and simulation yield an inverse
subthreshold slope $S$ close to 60\,mV/dec for sufficiently
negative $V_{\rm gs-Si}$ values, confirming the bulk-switching
phenomena in our devices. It is  noteworthy that as $V_{\rm
gs-Si}$ varies, the simulated curves $I_d(V_{\rm gs-Al})$ also
show good quantitative agreements with experimental data in terms
of $\Delta V_{\rm gs-Si}$. The ON-current decreases with
increasing $V_{\rm gs-Si}$ because the Schottky barriers at the
contact limit the maximum current through the device. The height
of the Schottky barrier is determined by the nanotube band gap and
the work function of the contact metal. It is expected that by
using nanotubes of larger diameters and choosing proper metals for
contact electrodes (e.g.\ Pd), the Schottky barriers can be
effectively lowered to achieve an ON-current as high as
25\,$\mu$A~\cite{javey_Nature2003}. The agreement between
experiment and simulation results is encouraging because it not
only marks the quality of the model used here, but also provides
compelling evidence for the operation concept of dual-gate CNFETs.

Since the regions A of our dual-gate CNFET resembles the extended
source and drain in a conventional MOSFET structure, the switching
speed of the dual-gate CNFET is mainly determined by the active
gate in region B. For high-frequency applications, this active
gate length $L_B$ of region B determines the transit time for an
electron/hole travelling from source to drain. Therefore, although
the entire length between the metal source/drain contacts is
$L=2L_A+L_B$, it is $L_B$, rather than $L$, that is the critical
parameter in terms of lateral scaling to improve performance. In
order to minimize parasitic capacitances for high-frequency
applications, we have also demonstrated that the Si back gate in
regions A can be replaced by using chemical dopants, although
future studies are needed to quantify the correlation between the
doping level and the obtained Fermi level shift.

\section{Conclusion}
In conclusion, we have fabricated CNFETs with a novel dual-gate
structure in order to achieve a distinct $p/i/p$ or $n/i/n$ doping
profile along the nanotube. The $p/i/p$ or $n/i/n$ doping scheme,
obtained by electrostatic and/or chemical doping effects,
eliminates the ambipolar characteristics and, for the first time,
creates pure $p$ or $n$-type enhancement-mode CNFETs with a
controllable polarity for thin gate oxides. Compared to previous
SB-CNFETs where the Schottky barriers dominate the switching
behavior, these dual-gate CNFETs are bulk-switched devices,
showing the steepest subthreshold slope reported to date. The good
agreement between experiment and simulation further corroborates
the novel device operation concept of the dual-gate structure.

\section*{Acknowledgement}
The authors thank Zhihong Chen, Marcus Freitag, and Jia Chen for
valuable discussions, and Bruce Ek for metal deposition and expert
technical assistance.
\newpage
\bibliographystyle{IEEEtran}

\begin{thebibliography}{10}
\bibitem{footnote1}
For example, the device studied in Ref.\,\cite{Javey_nanoLett04}
has used
  8-nm-thick HfO$_2$ ($\kappa\simeq 20$) and 10-nm-thick SiO$_2$ ($\kappa\simeq
  4$) layers as the gate insulator for top and back gates, respectively. In
  this geometry, the top gate efficiency is expected to be about 86\%.
\bibitem{foornote6}
We note that not all dual-gate CNFETs possess an inverse
subthreshold slope as
  steep as the one shown in Fig.\,\ref{variousVBG}, such as the device shown
  here. This could be due to the parasitic capacitance resulting from the
  unpassivated SiO$_2$ and/or Al$_2$O$_3$ surfaces.
\bibitem{footnote2}
In order to minimize computational time, the curves shown in
  Fig.\,\ref{exp_sim} were in fact simulated with gate lengths $L_A$ and $L_B$
  of 20\,nm for both regions A and B . The gate length of 20\,nm is found to be
  sufficient to describe a "long-channel" device, where the device
  characteristics does not depend on $L_A$ or $L_B$.
\end{thebibliography}

\newpage
\section*{Figures}
Fig.~1 (a) Schematics of a carbon nanotube field-effect transistor
(CNFET) with a back-gate configuration. (b) Measured subthreshold
characteristics $I_d(V_{\rm gs})$ of a typical CNFET (nanotube
diameter $\sim 1.4$\,nm) at different drain voltages. The gate
oxide consists of a layer of thermally-grown SiO$_2$ (10-nm
thick), and the contacts are made of Ti. The channel length
between the source and drain contacts is about 300\,nm.\\

Fig.~2 Schematic band diagrams of a CNFET for (a) $V_{\rm gs} < 0$
and (b) $V_{\rm gs}>0$. Two different drain voltages $V_{ds1}$ and
$V_{\rm ds2}$ ($V_{\rm ds2} < V_{\rm ds1} < 0$) are represented by
solid and
dash lines, respectively.\\

Fig.~3 Composite of the device layout of a dual-gate CNFET,
showing the SEM image of a CNFET with an Al middle gate underneath
the nanotube\cite{Lin_DRC2004}. The area between the Al gate and
the source/drain is denoted as regions A, and the Al gate is
denoted as region B.\\

Fig.~4 Measured drain current as a function of Al gate voltage
$V_{\rm gs-Al}$, with the Si back gate floating. \\

\begin{figure}
\center
\includegraphics[width=12cm]{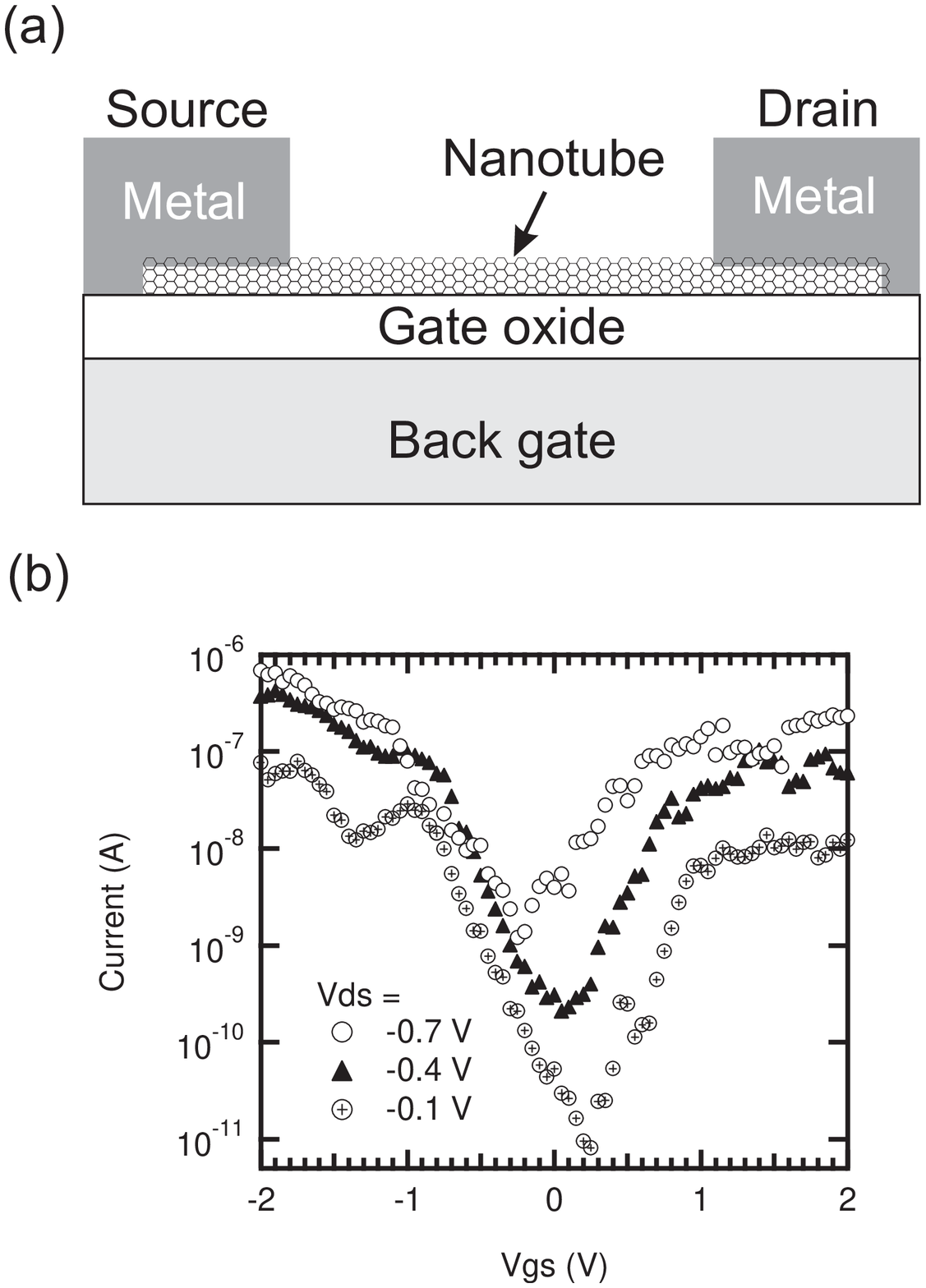}
\vspace{2em} \caption{Lin et al.} \label{thinoxCNFET}
\end{figure}

\begin{figure}
\center
\includegraphics[width=12cm]{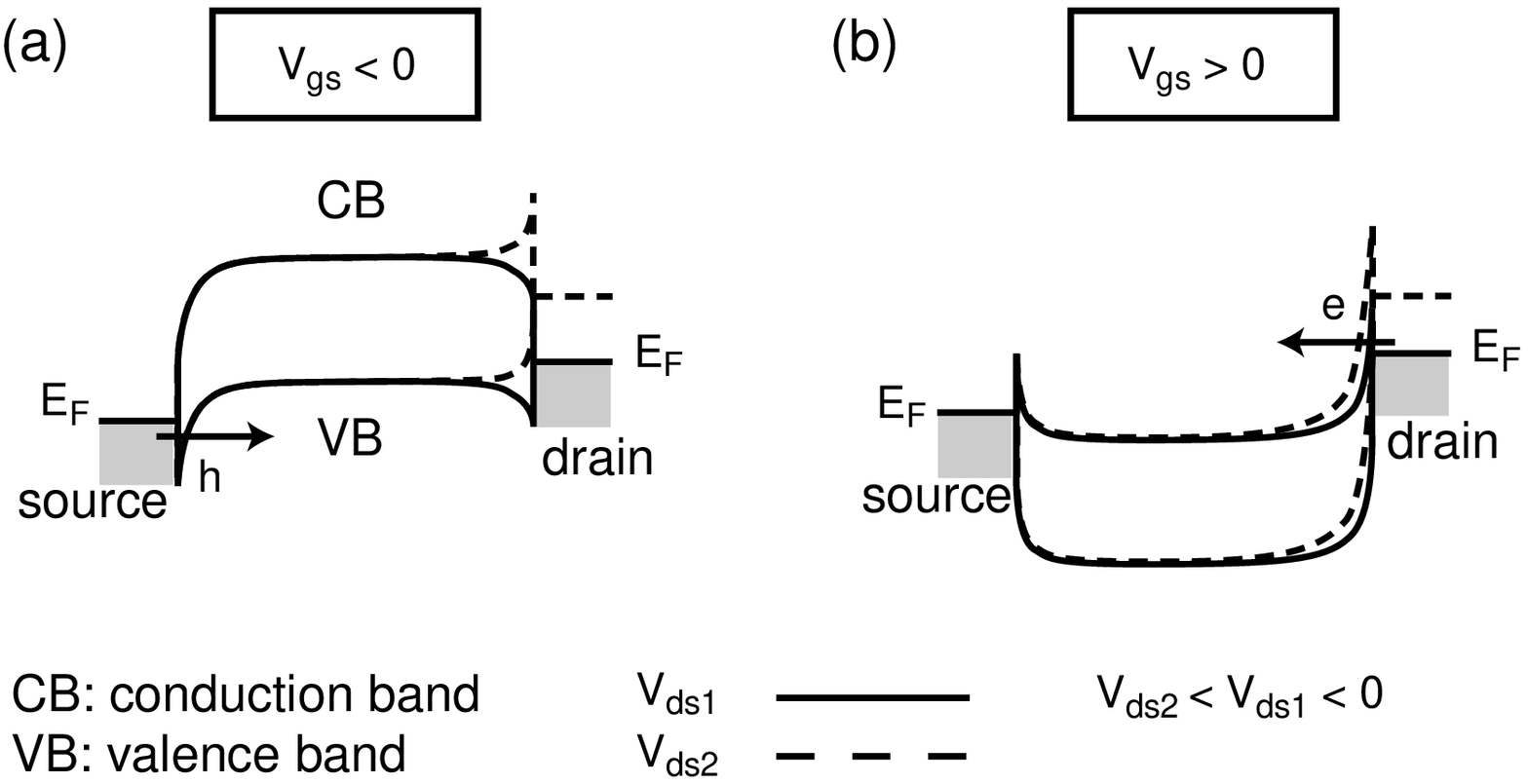}\vspace{2em}
\caption{Lin et al. } \label{bands2}
\end{figure}

\begin{figure} 
\center
\includegraphics[width=12cm]{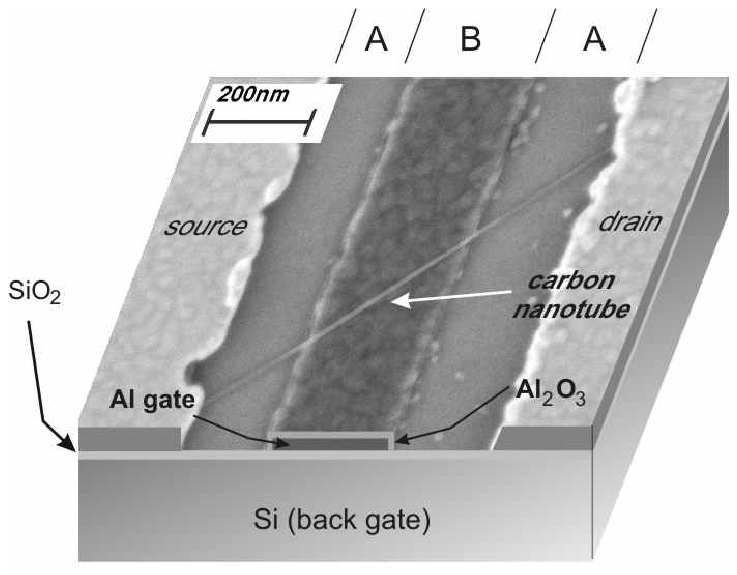}\vspace{2em}
\caption{Lin et al.} \label{dblGate}
\end{figure}

\begin{figure}
\center
\includegraphics[width=12cm]{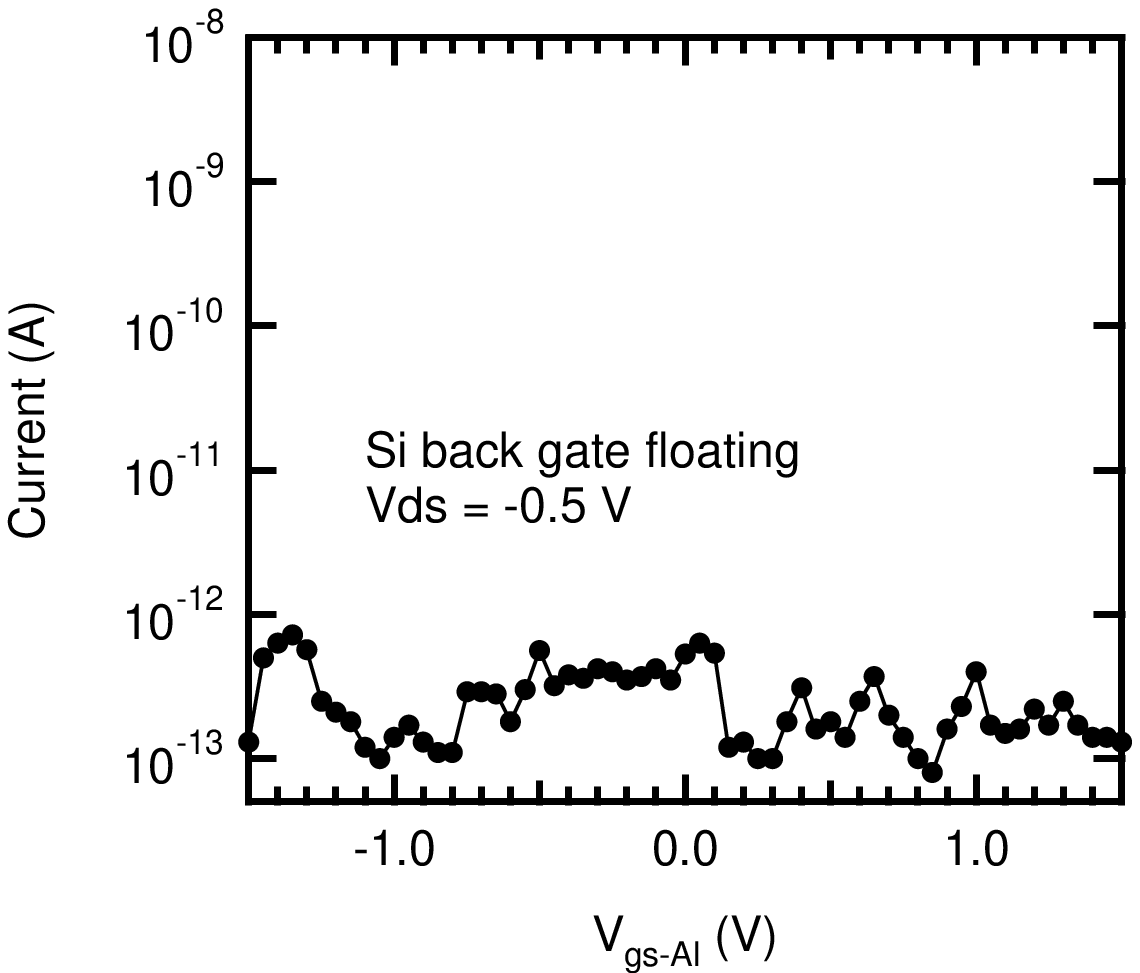}\vspace{2em}
\caption{Lin et al.} \label{BGfloating}
\end{figure}

\begin{figure}
\center
\includegraphics[width=12cm]{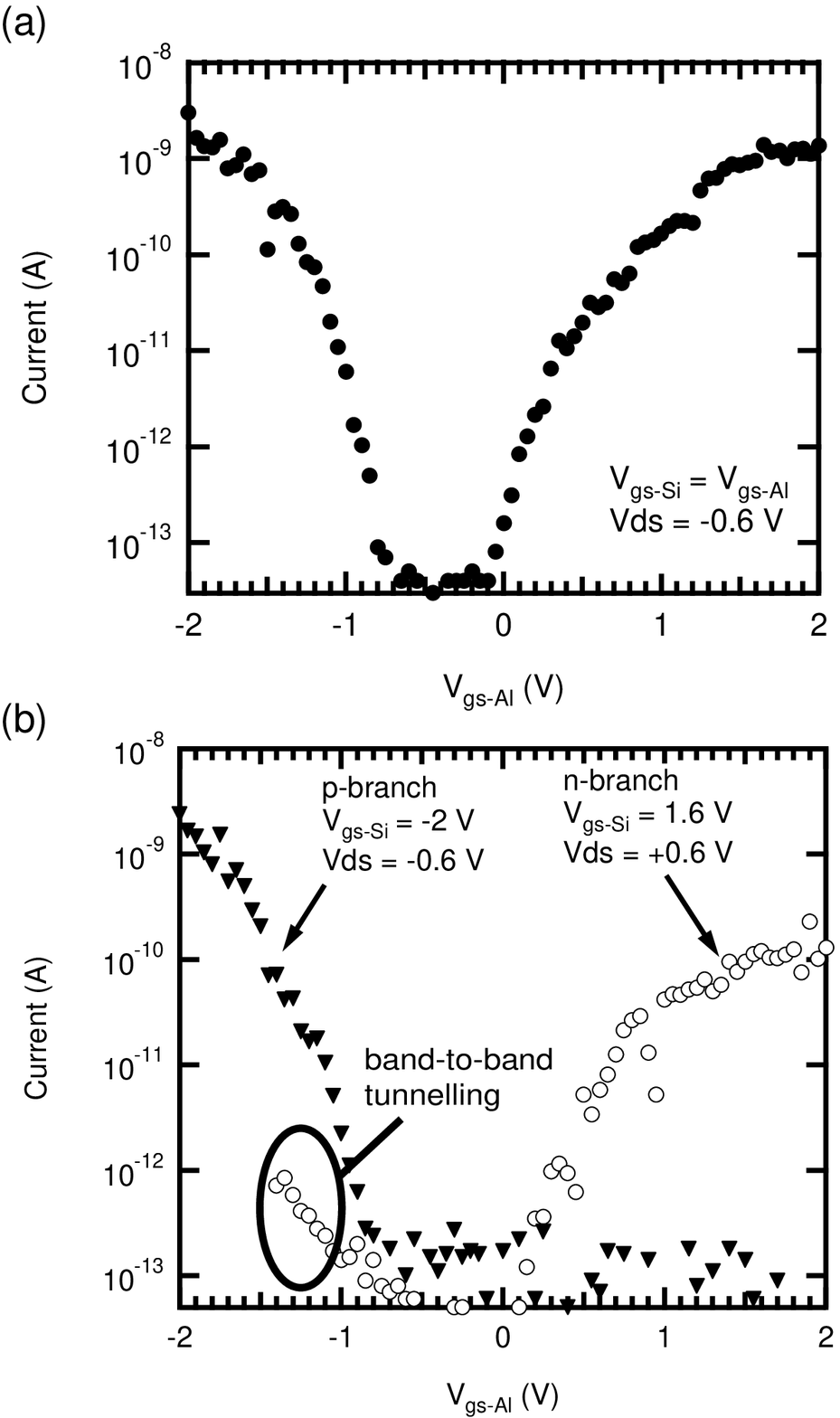}\vspace{2em}
\caption{Lin et al.} \label{npBranch}
\end{figure}

\begin{figure}
\center
\includegraphics[width=10cm]{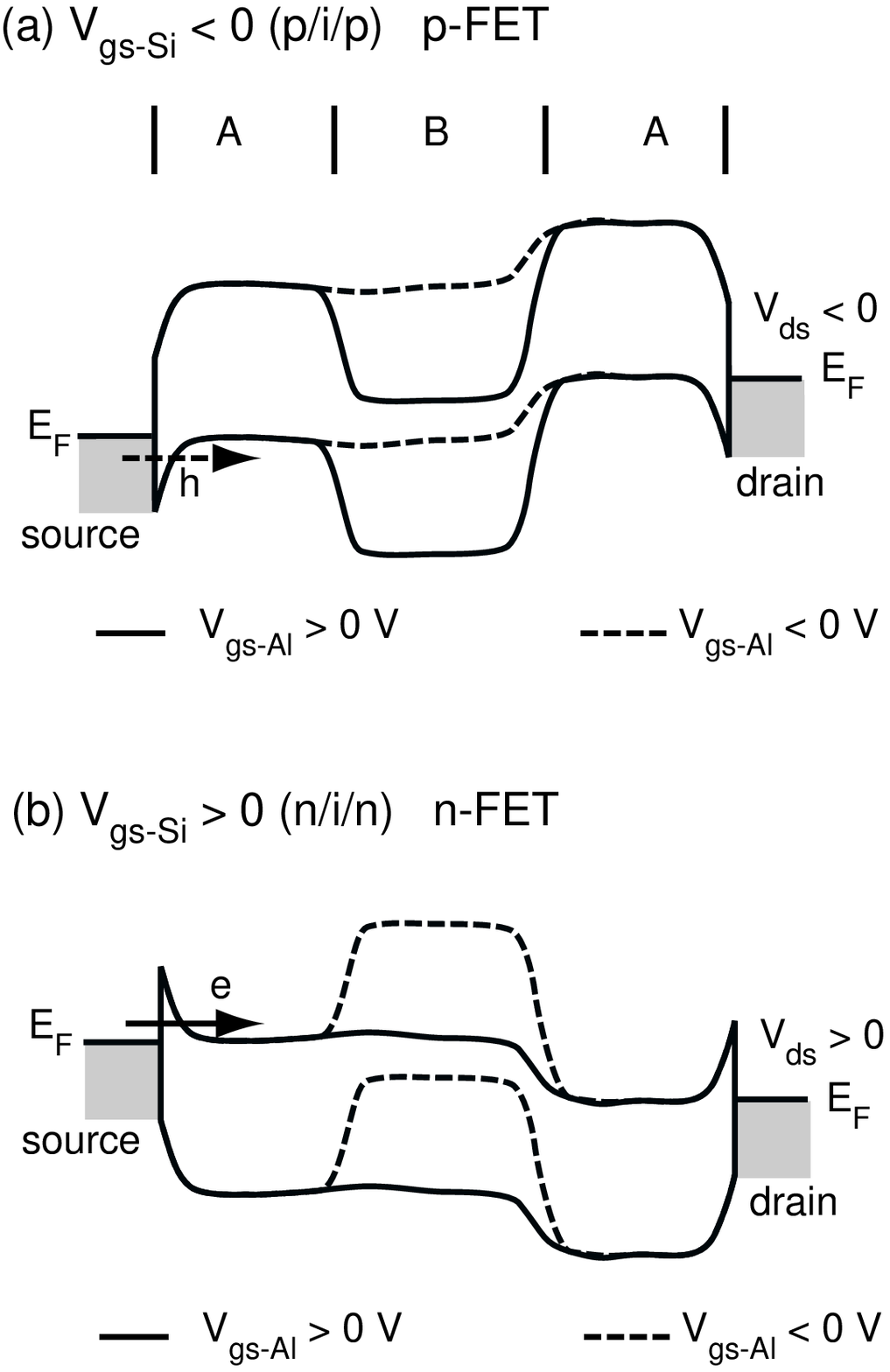}\vspace{2em}
\caption{Lin et al.} \label{bands3}
\end{figure}

Fig.~5 (a) Subthreshold characteristics of a dual-gate CNFET at
$V_{ds}=-0.6$\,V when the two gate voltages are set equal ($V_{\rm
gs-Si}=V_{\rm gs-Al}$), resembling a standard back-gated CNFET as
shown in Fig.\,\ref{thinoxCNFET}(a). (b) Subthreshold
characteristics ($I_d$ vs.\ $V_{\rm gs-Al}$) of the same dual-gate
CNFET measured at constant Si gate voltages $V_{\rm gs-Si}=-2$\,V
($\bullet$) and +1.6\,V ($\circ$), exhibiting clear $p$ and
$n$-type unipolar behaviors, respectively. Positive and negative
drain voltages are used for $n$-type and $p$-type branches,
respectively, in order to illustrate the operation as regular $n$
and $p$-FET devices.\\

Fig.~6 Schematic band diagrams of a dual-gate CNFET for (a)$V_{\rm
gs-Si}<0$ and (b)$V_{\rm gs-Si}>0$. In both (a) and (b), the solid
and the dashed lines represent situations for $V_{\rm gs-Al}> 0$
and $V_{\rm gs-Al}< 0$, respectively.\\

Fig.~7 (a) Measured subthreshold characteristics of a dual-gate
CNFET for different Si back gate voltages $V_{\rm gs-Si}$. (b)
measure $I_d$ vs. $V_{\rm gs-Al}$ when $V_{\rm gs-Si}=V_{\rm
gs-Al}$.\\

\begin{figure}
\center
\includegraphics[width=12cm]{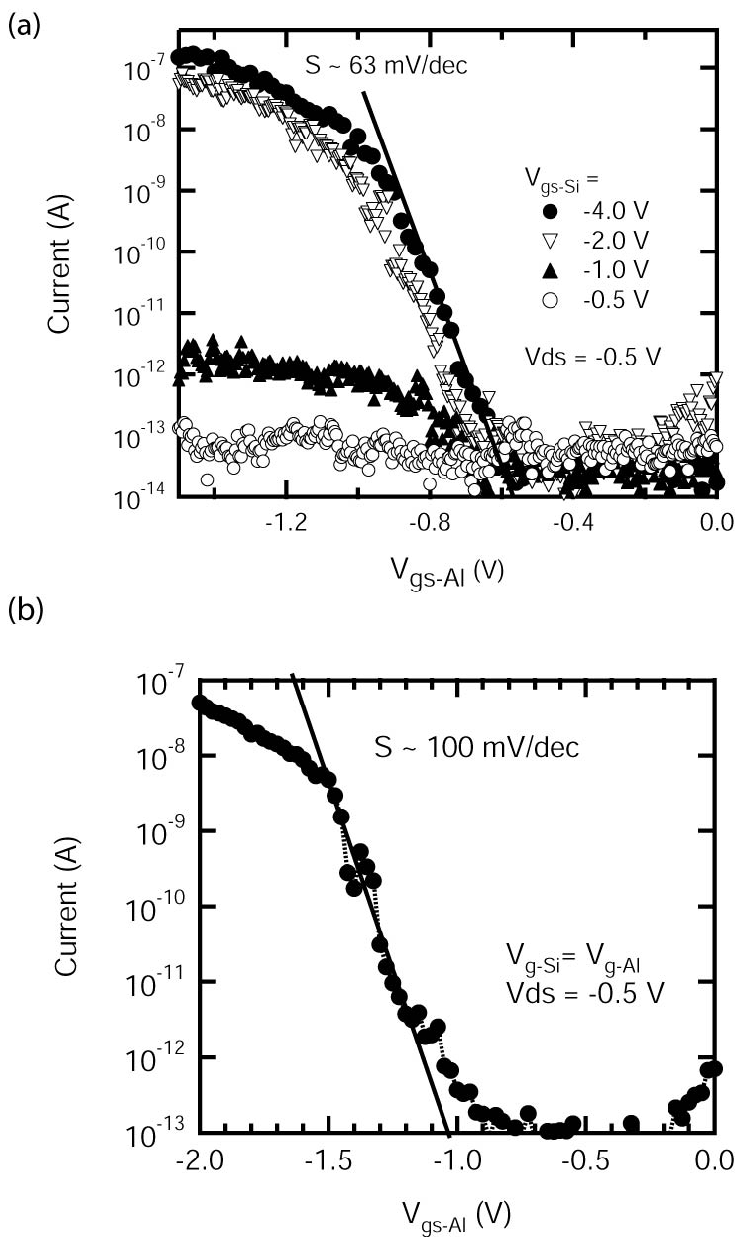}\vspace{2em}
\caption{Lin et al.} \label{variousVBG}
\end{figure}

\begin{figure}
\center
\includegraphics[width=12cm]{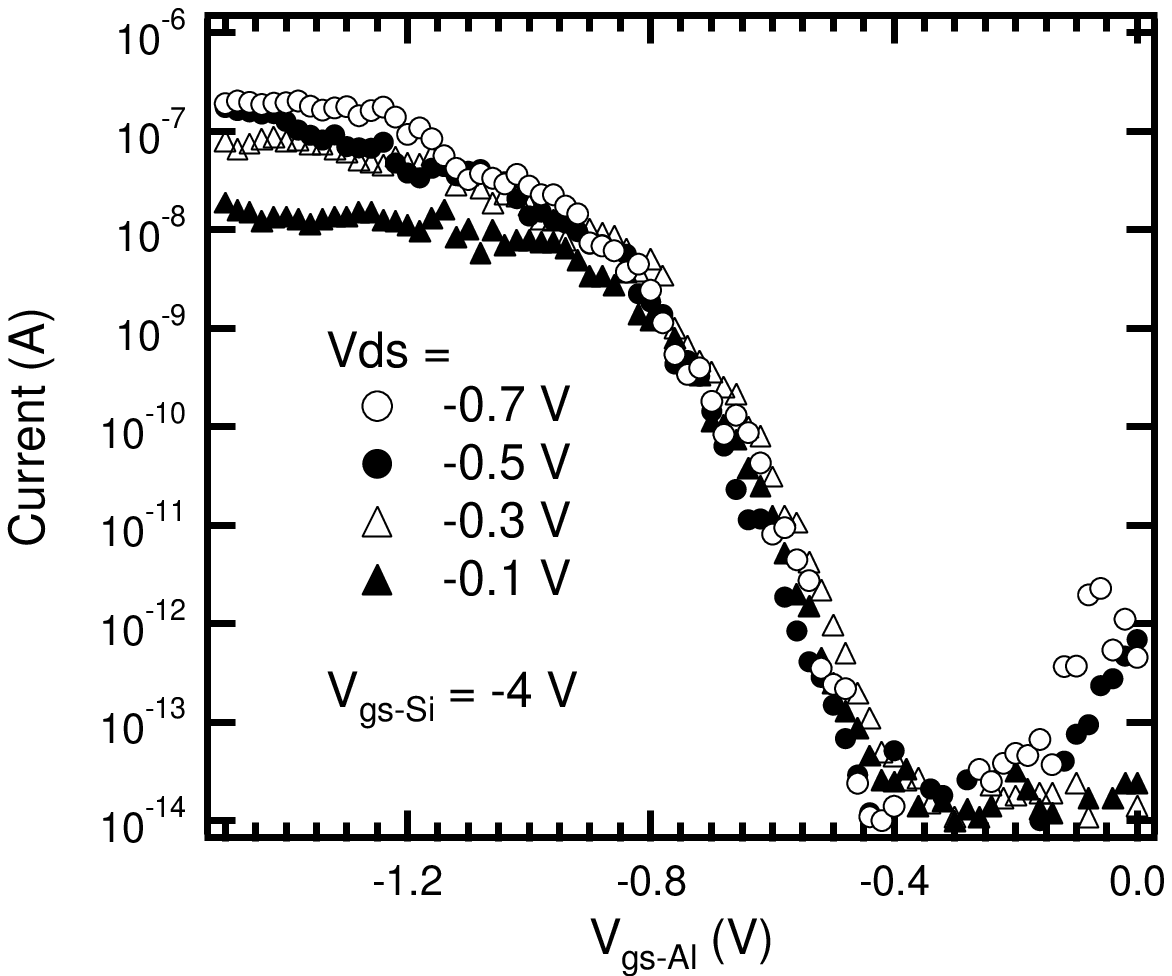}\vspace{2em}
\caption{Lin et al.} \label{variousVds}
\end{figure}

Fig.~8 Measured subthreshold characteristics of a dual-gate CNFET
(same device as the one in Fig.\,\ref{variousVBG}) for different
$V_{\rm ds}$ when the Si back gate $V_{\rm gs-Si}=-4$\,V\cite{Lin_DRC2004}.\\

Fig.~9 SEM image (top) and schematic cross-section diagram
(bottom) of a dual-gate CNFET with a layer of HSQ resist patterned
on top of the nanotube and the Al gate. The HSQ layer (100-nm
thick), denoted as region B', is used here in order to obtain a
chemical doping profile along the nanotube with respect to regions
A' and B'.\\

\begin{figure}
\center
\includegraphics[width=10cm]{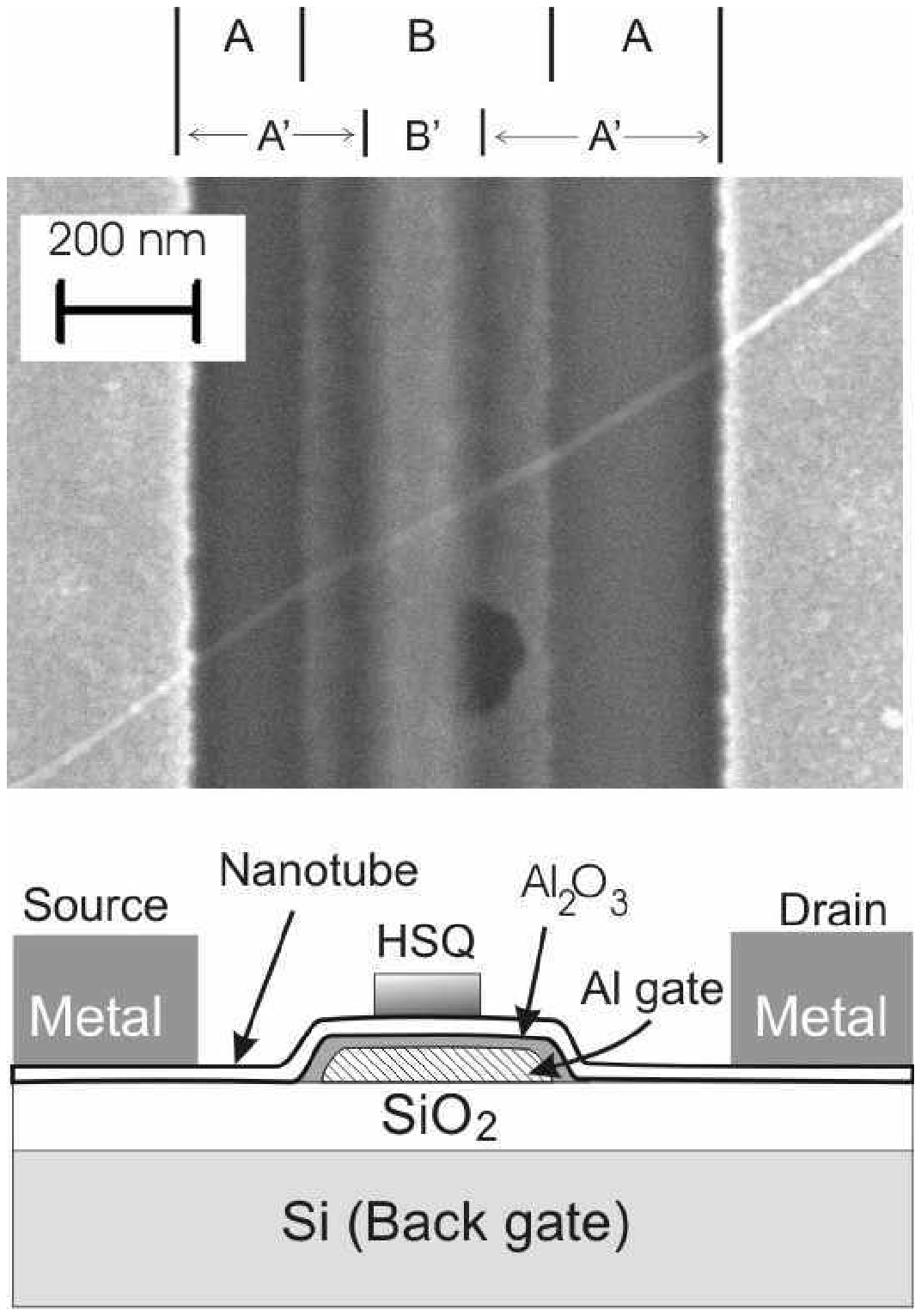}\vspace{2em}
\caption{Lin et al.} \label{dblGate_HSQ}
\end{figure}

\begin{figure} \center
\includegraphics[width=12cm]{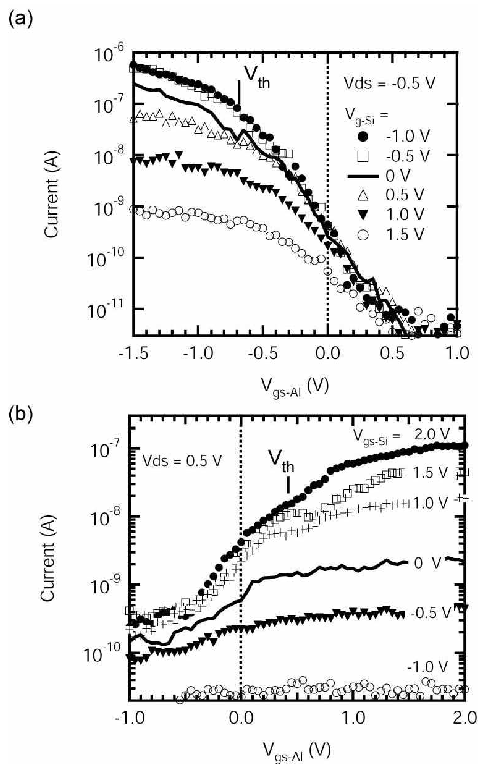}\vspace{2em}
\caption{Lin et al.} \label{p2n}
\end{figure}

\begin{figure}
\center
\includegraphics[width=12cm]{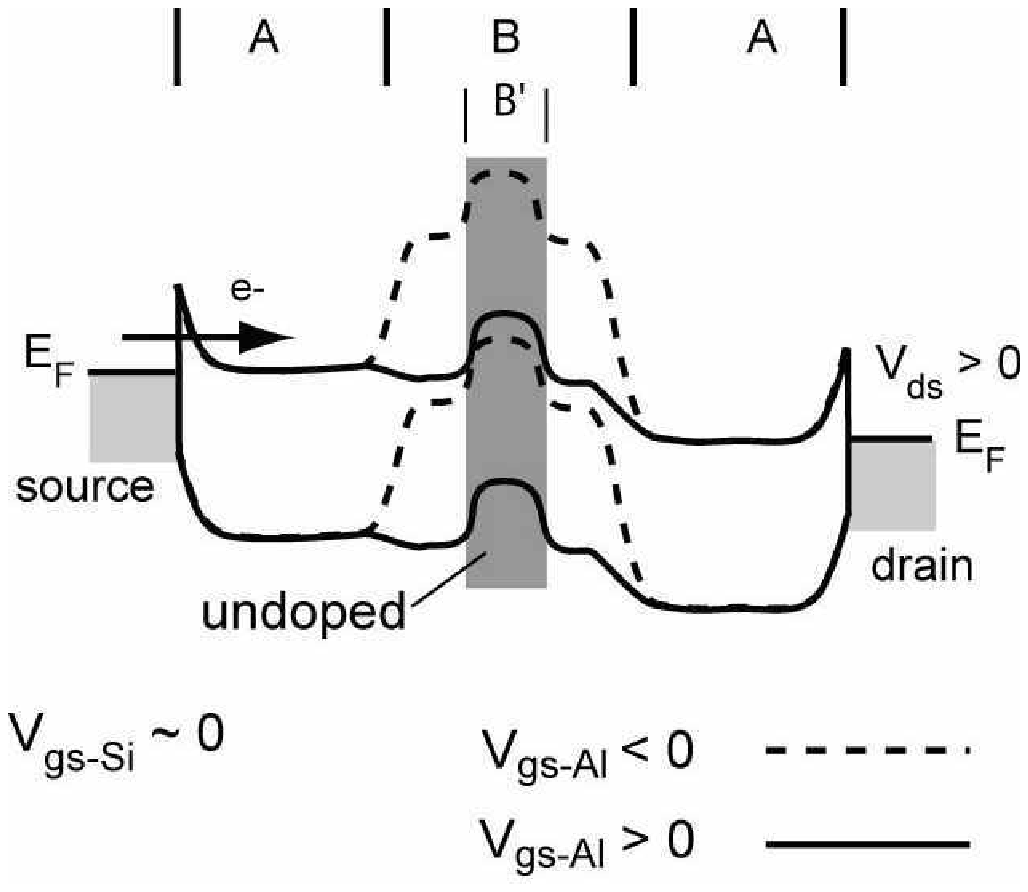}\vspace{2em}
\caption{Lin et al.} \label{doped_band3}
\end{figure}

\begin{figure} \center
\includegraphics[width=12cm]{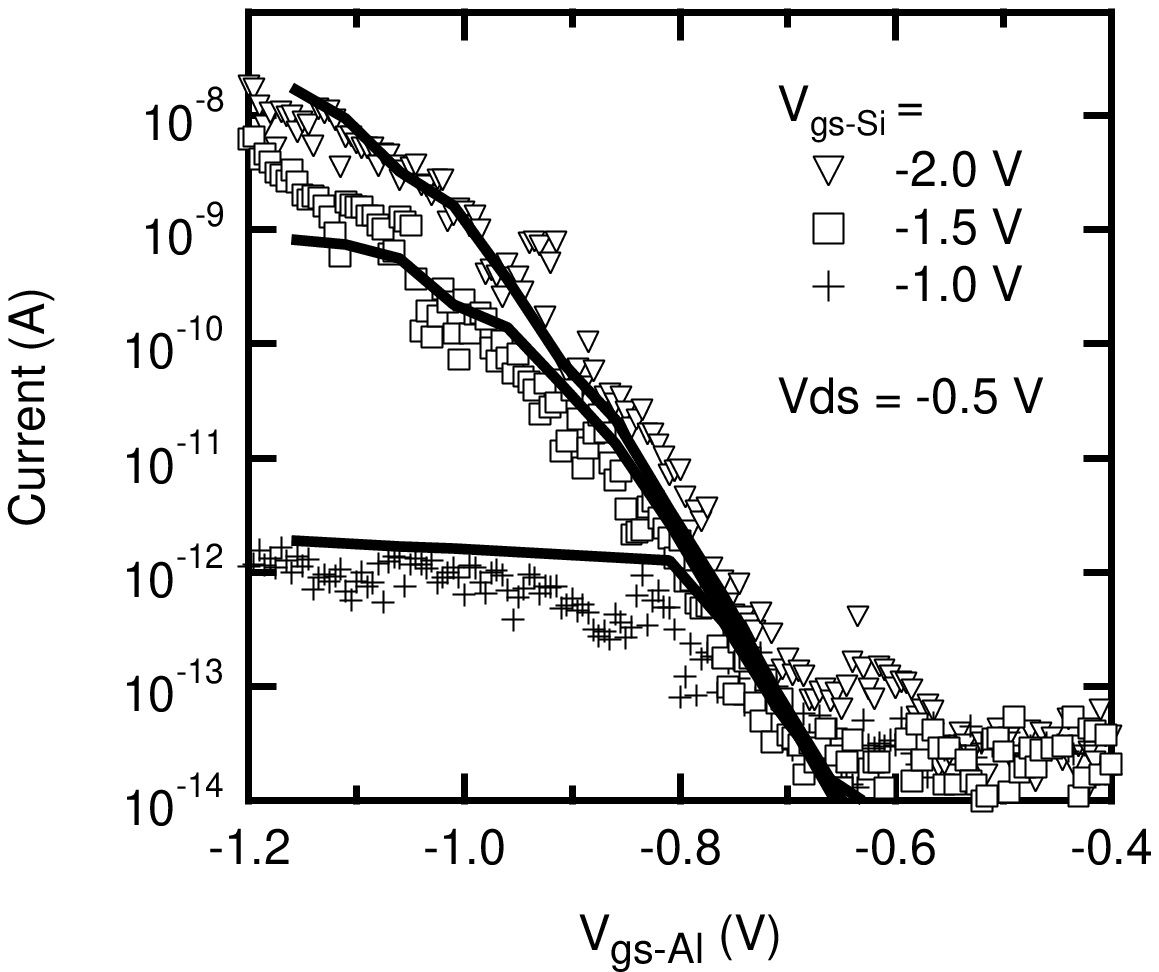}\vspace{2em}
\caption{Lin et al.} \label{exp_sim}
\end{figure}

Fig.~10 Measured subthreshold swing ($I_d$ vs.\ $V_{\rm gs-Al}$)
of the CNFET (Ti/Pd contact) device as shown in
Fig.\,\ref{dblGate_HSQ} for different Si back gate voltages (a)
before and (b) after PEI doping.\\

Fig.~11 Schematic band diagram of a dual-gate CNFET doped by
$n$-type dopants in regions A and part of region B. The shaded
area represents the undoped nanotube segment covered by the HSQ
resist (region B' in Fig.\,\ref{dblGate_HSQ}). Solid and dashed
lines represent the band bending for $V_{\rm gs-Al}>0$ and $V_{\rm
gs-Al}>0$, respectively.\\

Fig.~12 Subthreshold characteristics of a dual-gate CNFET (the
same device in Fig.\,\ref{variousVBG}) at different $V_{\rm
gs-Si}$. Simulated results are shown as solid curves for four Si
back gate voltages (from bottom to top): -0.15\,V, -0.5\,V, and
-1.0\,V. The simulated curves are translated by -0.66\,V along the
$x$-axis to compare with experimental data.\\

\clearpage
\clearpage

\clearpage

\clearpage

\clearpage

\clearpage

\clearpage

\clearpage

\clearpage

\clearpage

\clearpage

\clearpage

\end{document}